\documentclass[twoside,twocolumn,english,pra,twocolumn,aps,superscriptaddress,showpacs]{revtex4-1}

\usepackage[T1]{fontenc}
\usepackage[utf8x]{inputenc}
\usepackage{xcolor}
\usepackage{babel}
\usepackage{textcomp}
\usepackage{amsmath}
\usepackage{amsthm}
\usepackage{graphicx}
\usepackage[unicode=true,pdfusetitle,
 bookmarks=true,bookmarksnumbered=false,bookmarksopen=false,
 breaklinks=false,pdfborder={0 0 0},pdfborderstyle={},backref=false,colorlinks=true]
 {hyperref}

\makeatletter
\theoremstyle{plain}

\usepackage{tabularx}

\usepackage{makecell}
\usepackage{fontenc}

\hypersetup{urlcolor=blue}

\usepackage{cmap}
\usepackage[T1]{fontenc}

\input glyphtounicode
\makeatother

\begin{document}
\preprint{wThis line only printed with preprint option}
\title{Supersensitivity of Kerr phase estimation with two-mode squeezed vacuum
states}
\author{Yun-Feng Guo}
\affiliation{Institute of Quantum Information and Technology, Nanjing University
of Posts and Telecommunications, Nanjing 210003, China}
\author{Wei Zhong}
\email{zhongwei1118@gmail.com}

\affiliation{Institute of Quantum Information and Technology, Nanjing University
of Posts and Telecommunications, Nanjing 210003, China}
\affiliation{National Laboratory of Solid State Microstructures, Nanjing University,
Nanjing 210093, China}
\author{Lan Zhou}
\affiliation{School of Science, Nanjing University of Posts and Telecommunications,
Nanjing 210003, China}
\author{Yu-Bo Sheng}
\affiliation{College of Electronic and Optical Engineering, Nanjing University
of Posts and Telecommunications, 210003, Nanjing, China}
\begin{abstract}
We analytically investigate the sensitivity of Kerr nonlinear phase
estimation in a Mach-Zehnder interferometer with two-mode squeezed
vacuum states. We find that such a metrological scheme could access
a sensitivity scaling over the Boixo \emph{et al.}'s generalized sensitivity
limit {[}S. Boixo \emph{et al}., Phys. Rev. Lett. \textbf{98}, 090401
(2007){]}, which is saturable with celebrated NOON states. We also
show that parity detection is a quasioptimal measurement which can
nearly saturate the quantum Cram\'er-Rao bound in the aforementioned
situation. Moreover, we further clarify the supersensitive performance
observed in the above scheme is due to the restriction of Boixo \emph{et
al}.'s generalized sensitivity limit (BGSL) to probe states with fixed
photon numbers. To conquer this problem, we generalize the BGSL into
the case with probe states of a fluctuating number of photons, to
which our scheme belongs.
\end{abstract}
\maketitle

\section{Introduction}

The interferometer is a widely used optical device allowing one to
implement precision measurements ranging from the first measurements
of the speed of light to modern microscopic imaging and gravitational
wave detection \citep{Giovannetti2011,Pirandola2018NPreview,Lawrie2019ACSPhoto,Polino2020review}.
The optical interferometer has an atomic analog \citep{Pezze2018RMP},
too. Based on these devices, the problem of measurement of an unknown
physical quantity is converted to the problem of estimating the relative
phase shift between the two modes of the interferometer. Hence the
sensitivity of phase estimation is a crucial factor to determine the
performance of specific applications of precision measurement. 

For linear phase estimations, the optical interferometer with $N$
uncorrelated photons is highly possible to attain a phase uncertainty
scaling as $N^{-1/2}$, as a consequence of the quantum fluctuation
of photons. This sensitivity scaling is also referred to as the shot-noise
limit, which would be broken when quantum resources are taken into
account. To obtain sub-shot-noise-limit sensitivities, using non-classical
states of light has been theoretically and experimentally confirmed
as an effective way. A large number of non-classical states have been
proposed to enhance the estimation sensitivity in both optical and
atomic interferometry \citep{Pezze2018RMP,Lawrie2019ACSPhoto}, such
as squeezed states \citep{Caves1981PRD,Pezze2008PRL,Anisimov2010PRL,Gross2011Nature,Hamley2012NatPhys,Lang2013PRL,Birrittella2014JOSAB,Braun2014PRA,Peise2015NC,Kruse2016PRL,Barsotti2018RP,Ataman2019PRA,Ataman2020PRA,Zhong2020},
Fock states \citep{Holland1993PRL,Campos2003,Lucke2011Science,Pezze2013PRL},
entangled coherent (EC) states, \citep{Joo2011PRL} and other robust
quantum states \citep{Huver2008PRA,Dorner2009PRL,Demkowicz-Dobrzanski2009PRA,Jiang2012PRA,Zhong2017PRA,Zhong2021PRA},
etc. Moreover, using NOON states is expected to attain the Heisenberg
limit  $N^{-1}$, which was known as the ultimate accessible sensitivity
in the linear phase estimation \citep{Holland1993PRL,Dowling2008Review,Zwierz2010PRL}. 

In a seminal work \citep{Boixo2007PRL}, Boixo \emph{et al}. developed
generalized sensitivity limits scaling with $N^{-k}$ for single-parameter
estimation with $k$ the nonlinearity order of the Hamiltonian governing
the system dynamics and $N$ the total number of particles in the
system. By replacing $k$-body interactions by $N$-body interactions,
Roy \emph{et al}. showed that an exponential enhanced accuracy would
be obtained \citep{Roy2008PRL}. According to the Boixo \textit{et
al}.'s generalized sensitivity limits (BGSLs), the ultimate sensitivity
for a second-order nonlinear phase estimation (i.e., $k=2$) should
scale as $N^{-2}$, a $N$ factor improvement over the Heisenberg
limit (see Sec.~IV for detailed discussions). Due to its potential
to suppress the conventional Heisenberg limit in contrast to the linear
one \citep{Hall2012PRX}, nonlinear phase estimation has been receiving
increasing attention. Several works identified that a Heisenberg-limit-scaling
sensitivity is attainable even without the use of entangled resources
in the nonlinear cases with single and two field modes \citep{Luis2004PLA,Beltran2005PRA,Boixo2008PRL,Rivas2010PRL,Napolitano2011Nat,Chen2018NC}.
Recently, It has shown that using coherent state of light may provide
better accuracy than the Heisenberg limit in Kerr phase estimation
by a scaling factor of $\bar{N}^{-3/2}$ with $\bar{N}$ the total
mean number of photons \citep{Woolley2008NJP,Zhang2019PRA}. Moreover,
Joo \emph{et al}. demonstrated that, EC states of small photon numbers
can outperform NOON states in nonlinear settings, suggesting that
the BGSL would be further overcome \citep{Joo2012PRA}. Besides the
demonstrations in optical systems, the study of nonlinear metrology
in the atomic area is growing rapidly, such as interaction-based measurement
of ensemble magnetization \citep{Napolitano2011Nat,Napolitano2010aNJP},
precision measurement of atomic scattering \citep{Rey2007PRA,Boixo2008PRL,Choi2008PRA},
etc. 

\begin{figure}[t]
\centering{}\includegraphics{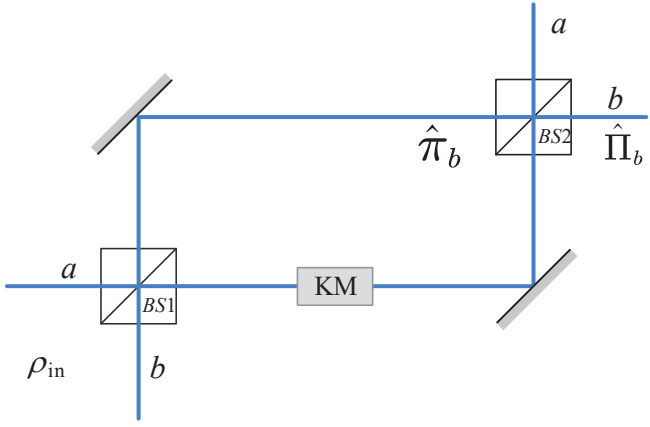}\caption{(Color online) Sketch of the Kerr nonlinear phase estimation in a
standard Mach-Zehnder interferometer with a single-port parity detection.
\label{fig:MZI}}
\end{figure}

In this manuscript, we address the problem of Kerr-medium-induced
phase estimation in a two-mode optical interferometry with two-mode
squeezed vacuum (TMSV) states (see Fig.~\ref{fig:MZI}), due to its
high feasibility in both optical \citep{Weedbrook2012RMP,Pirandola2018NPreview}
and atomic \citep{Gross2011Nature,Hamley2012NatPhys,Peise2015NC,Kruse2016PRL,Pezze2018RMP}
experiments. By invoking the phase averaged approach, we analytically
derive the ultimate phase uncertainties set by quantum Cram\'er-Rao
bound in the aforementioned situation. We also identify that parity
detection is a nearly optimal measurement for saturating the phase
uncertainties we derived. Our results suggest that the scheme with
TMSV states is highly superior to the one with EC states \citep{Joo2012PRA},
and also acquires a sensitivity beyond the BGSL \citep{Boixo2007PRL}.
More importantly, this sensitivity superiority effect can still be
observed when the photon number becomes large, which is in sharp contrast
with the previous result reported in Ref.~\citep{Joo2012PRA}. Moreover,
although widely used in various applications related to quantum teleportation
\citep{Liuzzo-Scorpo2017PRL}, quantum dense coding \citep{Ban1999JOB},
quantum illumination \citep{Tan2008PRL,Sanz2017PRL,Jo2021PRR}, quantum
state tomography based on ultracold atomic ensemble \citep{Peise2015NC},
and especially linear phase estimation \citep{Anisimov2010PRL,Carranza2012JOSAB,Kruse2016PRL,Ouyang2016JOSAB,Zhang2017oe},
the TMSV states have never been discussed in the literature on Kerr
nonlinear phase estimation. Thus our work also serves to complement
studies in this aspect.

This paper is organized as follows. In Sec. II, we introduce the Kerr
phase estimation setup and derive the accessible phase uncertainties
for both twin Fock (TF) and TMSV states with a single-port parity
detection scheme. In Sec. III, we analytically compute the ultimate
sensitivities for the above two states via the quantum Fisher information
(QFI). In Sec. IV, we revisit the derivation of the BGSL and generalize
it to the cases with probe states with variable photon numbers. Finally,
our conclusions are given in Sec. V.

\section{Achievable sensitivities with parity detection }

The scheme of our Mach-Zehnder interferometer (MZI) setup is depicted
in Fig.~\ref{fig:MZI}. A MZI is usually composed of two balanced
beam splitters $B_{i}\;\left(i=1,2\right)$ and a phase shifting $U_{\varphi}$
with $\varphi$ to be estimated. During the photon propagation between
the BSs, an unknown phase of interest is accumulated. According to
different propagation mechanisms, the operation of a phase shifter
acting on the lower mode can be formally modeled by 
\begin{equation}
U_{\varphi}=\exp[-i\varphi(a^{\dagger}a)^{k}],\label{eq:phaseshift}
\end{equation}
where $a^{\dagger}\left(a\right)$ stands for the creation (annihilation)
operator of the corresponding mode and the exponent $k$ denotes the
order of nonlinearity. In this expression, $k=1$ corresponds to the
linear phase shift and $k=2$ to the Kerr nonlinear phase shift \citep{Luis2004PLA,Beltran2005PRA}.
Physically, they may describe the behavior of light propagating in
free space and Kerr medium, respectively. 

In what follows, we mainly focus on the case of $k=2$, i.e., a Kerr
phase shifting. Given $B_{i}\;\left(i=1,2\right)$ (see below for
specific expressions) and $U_{\varphi}$, the dynamics of the Mach-Zehnder
interferometer is represented as a compound operation, i.e., $K_{\varphi}=B_{2}\,U_{\varphi}B_{1}$.
Let $\rho_{{\rm in}}$ denote the state of light entering at the input
ports of the interferometer. Then, the state at the output ports reads
$\rho_{{\rm out}}=K_{\varphi}\rho_{{\rm in}}K_{\varphi}^{\dagger}.$
 Finally, a measurement is performed at the output port of the interferometer
and then the true value of the phase is extracted from the measurement
outcomes. Given a measurement observable $\mathcal{O}$, the value
of $\varphi$ can be inferred from the average value of the observable
$\langle\mathcal{O}\rangle$. The real accessible precision on $\varphi$
is given by the error-propagation formula as follows \citep{Yurke1986PRA}:

\begin{eqnarray}
\Delta\varphi & = & \frac{1}{\sqrt{\upsilon}}\frac{\sqrt{\langle\mathcal{O}^{2}\rangle-\langle\mathcal{O}\rangle^{2}}}{\left|\partial\langle\mathcal{O}\rangle/\partial\varphi\right|},\label{eq:error_propagation_formula}
\end{eqnarray}
with $\langle\cdot\rangle\equiv{\rm Tr}\left(\cdot\;\rho_{{\rm out}}\right)$
being the expectation and $\upsilon$ the repetitions of the experiment.
One should note that applying this method requires $\left\langle \mathcal{O}\right\rangle $
to be a monotonous function of the parameter $\varphi$ at least in
a local region of parameter values determined from prior knowledge
\citep{Pezze2018RMP}.

We assume the TMSV states of light as the input states of the interferometer
in the above setting. The TMSV states can be understood as a linear
superposition of TF states $\vert n,n\rangle$ \citep{Lucke2011Science}
(which is known as the Holland-Burnett state in the optical setting
\citep{Holland1993PRL}) as

\begin{equation}
\vert\psi_{{\rm TMSV}}\rangle=\sum_{n=0}^{\infty}\sqrt{p_{n}}\vert n,n\rangle,\label{eq:TMSVS}
\end{equation}
where 
\begin{equation}
p_{n}=\left(1-\frac{\bar{N}}{\bar{N}+2}\right)\left(\frac{\bar{N}}{\bar{N}+2}\right)^{n},\label{eq:weights}
\end{equation}
with $\bar{N}$ the average photon number \citep{Gerry2004Book}.
Hence the output state is given by $\vert\psi_{{\rm out}}\rangle=K_{\varphi}\vert\psi_{{\rm TMSV}}\rangle$.
As shown in Fig.~\ref{fig:MZI}, a single-port parity detection is
assumed to be carried out on the output mode $b$. The parity measurement
was originally proposed to probe atomic frequency in trapped ions
by Bollinger \textit{et al.} \citep{Bollinger1996PRA} and later employed
for optical interferometry by Gerry \citep{Gerry2000}. It accounts
for distinguishing the states with even and odd numbers of photons
in a given output port. Specifically, the parity is assigned as the
value of +1 when the photon number of a state is even, and the value
of −1 if odd. Hence it can be formulated as
\begin{equation}
\Pi_{b}=\left(-1\right)^{b^{\dagger}b}={\rm exp}\left(i\pi b^{\dagger}b\right).
\end{equation}
Due to the identity $\Pi_{b}^{2}=\openone$ with $\openone$ being
the identity matrix, the calculation of sensitivity from Eq.~\eqref{eq:error_propagation_formula}
can be simplified to the calculation of the expectation value of $\Pi_{b}$
for the output state \citep{Seshadreesan2013PRA,Zhong2021PRA}. With
Eq.~\eqref{eq:TMSVS}, this expectation can be expressed as 
\begin{eqnarray}
\langle\Pi_{b}\rangle_{{\rm TMSV}} & = & \sum_{n=0}^{\infty}p_{n}\langle\Pi_{b}\rangle_{{\rm TF}}.\label{eq:information_tmsv}
\end{eqnarray}
where 
\begin{equation}
\langle\Pi_{b}\rangle_{{\rm TF}}\equiv\langle n,n\vert K_{\varphi}^{\dagger}\Pi_{b}K_{\varphi}\vert n,n\rangle\label{eq:expectation_TF}
\end{equation}
is the expectation value of $\Pi_{b}$ for TF states. Note that the
expression of Eq.~\eqref{eq:information_tmsv} is a direct consequence
of the commutation relation of the total photon number operator $a^{\dagger}a+b^{\dagger}b$
and the compound operator $K^{\dagger}\Pi_{b}K_{\varphi}$, i.e.,
$\left[a^{\dagger}a+b^{\dagger}b,K^{\dagger}\Pi_{b}K_{\varphi}\right]=0$,
which leads to a vanishing value of $\langle n^{\prime},n^{\prime}\vert K^{\dagger}\Pi_{b}K_{\varphi}\vert n,n\rangle$
when $n^{\prime}\neq n$. To analytically derive Eq.~\eqref{eq:expectation_TF},
we here consider the second BS operation as a part of measurement
and hence the parity measurement through the BS is transformed into
\citep{Chiruvelli2011,Jiang2012PRA}

\begin{equation}
\pi_{b}=B_{2}\Pi_{b}B_{2}^{\dagger}=\sum_{N=0}^{\infty}i^{N}\sum_{l=0}^{N}(-1)^{l}\vert l,N-l\rangle\langle N-l,l\vert,\label{eq:parity_BS}
\end{equation}
where the operation of $B_{2}$ is formulated by $B_{2}={\rm exp}\left[-i\pi\left(a^{\dagger}b+ab^{\dagger}\right)/4\right]$
following the form adopted in \citep{Campos2003}. It means that the
measurement on the output mode $b$ is equivalent to performing a
projective measurement $\pi_{b}$ to the state before the second BS
$B_{2}$, i.e., 
\begin{equation}
\left|\psi_{2n}\left(\varphi\right)\right\rangle \equiv U_{\varphi}B_{1}\vert n,n\rangle,\label{eq:TF_parametric}
\end{equation}
such that 
\begin{equation}
\langle\Pi_{b}\rangle_{{\rm TF}}=\langle\psi_{2n}\left(\varphi\right)\vert\pi_{b}\vert\psi_{2n}\left(\varphi\right)\rangle.
\end{equation}
In the Schrödinger representation, the parametric TF states of Eq.~\eqref{eq:TF_parametric}\textbf{
}can be explicitly expressed as follows:
\begin{eqnarray}
\left|\psi_{2n}\left(\varphi\right)\right\rangle  & = & \sum_{k=0}^{n}C_{nk}{\rm exp}\left(-i4k^{2}\varphi\right)\vert2k,2n-2k\rangle,\label{eq:tf_state_through_bs}
\end{eqnarray}
with
\begin{eqnarray}
C_{nk} & = & \left(-1\right)^{n-k}\frac{1}{2^{n}}\left[\binom{2k}{k}\binom{2n-2k}{n-k}\right]^{1/2}.\label{eq:conffience}
\end{eqnarray}
Here we simply select the first BS operation in the form of $B_{1}={\rm exp}\left[\pi\left(a^{\dagger}b-a\,b^{\dagger}\right)/4\right]$
as in \citep{Campos2003}. Although it does not satisfy the symmetric
relation $\hat{B}_{1}=\hat{B}_{2}^{\dagger}$ as usually assumed in
previous studies \citep{Jarzyna2012PRA,Zhong2017PRA}, the final measurement
results remain invariant, apart from a translation of $\pi/2$ in
terms of $\varphi$. With Eqs.~\eqref{eq:parity_BS}, \eqref{eq:tf_state_through_bs}
, and \eqref{eq:conffience}, we explicitly derive the expectation
of the parity operator for the TF states as 

\begin{eqnarray}
\langle\Pi_{b}\rangle_{{\rm TF}} & = & \sum_{k=0}^{n}C_{nk}^{2}{\rm cos}\left[4n(n-2k)\varphi\right].\label{eq:signal_tf}
\end{eqnarray}
Inserting Eq.~\eqref{eq:signal_tf} into Eq.~\eqref{eq:information_tmsv}
then finally yields the signal of parity measurement $\langle\Pi_{b}\rangle_{{\rm TMSV}}$
with respect to the TMSV states.

\begin{figure}[t]
\begin{centering}
\includegraphics[scale=0.95]{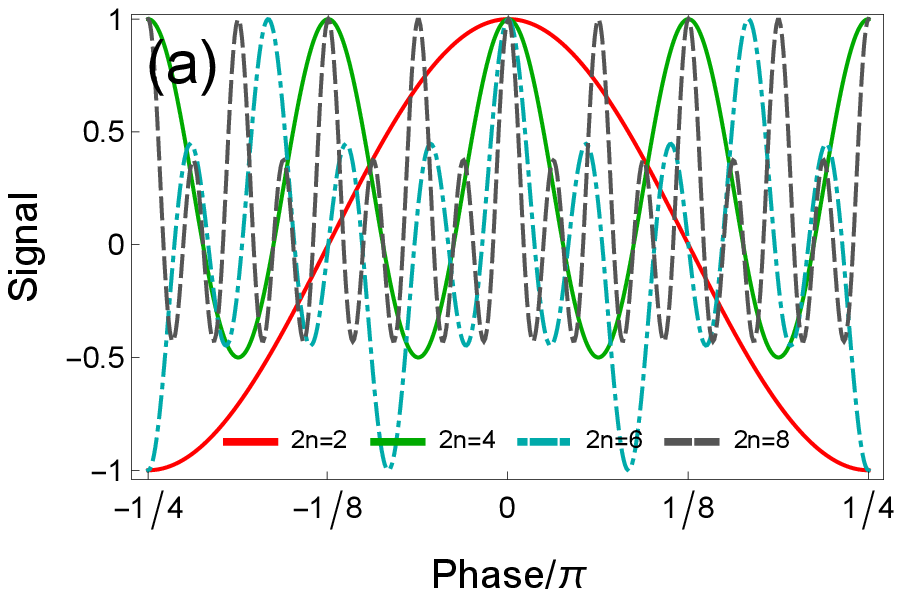}
\par\end{centering}
\centering{}\includegraphics[scale=0.95]{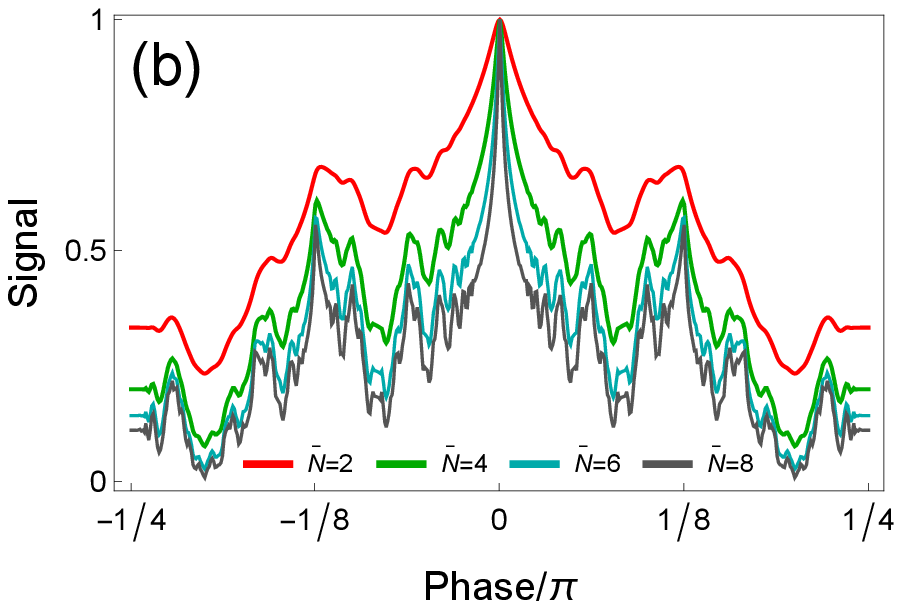}\caption{(Color online) Signals with parity detection versus phase for (a)
TF states $\vert\psi_{{\rm TF}}\rangle\equiv\vert n,n\rangle$ of
total photon number $2n$ and (b) TMSV states $\vert\psi_{{\rm TMSV}}\rangle$
of mean photon number $\bar{N}$. Different colors here refer to different
numbers of $2n$ and $\bar{N}$, respectively. In (a) the frequency
of signal oscillation becomes higher as $n$ increases. In (b) the
width of the peak at the zero phase point becomes narrower as $\bar{N}$
increases. \label{fig:signal}}
\end{figure}

We plot in Fig.~\ref{fig:signal} the signals with parity measurement
as a function of $\varphi$ for both TF and TMSV states according
to Eqs.~\eqref{eq:signal_tf} and \eqref{eq:information_tmsv}. At
first glance, they exhibit a completely different behavior. As seen
in Fig.~\ref{fig:signal}(a), the signal for the TF states has an
oscillation with a period depending on the total photon number $2n$.
In the cases with odd $n$, the oscillation amplitude varies within
the range between $-1$ and $1$ and, in the cases with even $n$,
it varies within the range between $-0.5$ and $1$. While, as shown
in Fig.~\ref{fig:signal}(b), the signal for the TMSV states does
not have a behavior of the periodic oscillation as presented in Fig.~\ref{fig:signal}(a)
and varies only within the range between $0$ and $1$. Another key
difference is that the signal for the TMSV states changes with the
period of $\pi/2$ rad irrespective of the average photon number $\bar{N}$.
In this case, the signal has a sharp peak at $\varphi=0$ and the
peak width becomes narrower as $\bar{N}$ increases, which will render
a significant improvement in sensitivity as shown in Fig.~\ref{fig:sensitivity}.
These distinctions can be understood from the expression of Eq.~\eqref{eq:information_tmsv},
which shows $\langle\Pi_{b}\rangle_{{\rm TMSV}}$ is the weighted
sum of $\langle\Pi_{b}\rangle_{{\rm TF}}$ with the weights $p_{n}$
given by Eq.~\eqref{eq:weights}. 

Furthermore, with Eqs.~\eqref{eq:signal_tf} and \eqref{eq:information_tmsv}
and according to Eq.~\eqref{eq:error_propagation_formula}, we numerically
plot in Fig.~\ref{fig:sensitivity} the phase uncertainties of Kerr
phase measurement with parity detection for the TF and TMSV states
around the zero-point of $\varphi$. As a contrast, we take the EC
state as a benchmark with the same detection strategy (see Appendix
A for detailed derivation), and plot in Fig.~\ref{fig:sensitivity}
the phase uncertainty corresponding to Eq.~\eqref{eq:error-ECS}.
Our results indicate that the TF state asymptotically approaches the
BGSL $\bar{N}^{-2}$ as the number of photons decreases and saturates
the limit at $\bar{N}=2$ (see Sec.~IV for a demonstration of saturation
of the BGSL), where the probe state is a NOON state as a result of
the Hong-Ou-Mandel effect \citep{Hong1987PRL}. We see that the TMSV
state has a significant improvement in sensitivity over the EC state
and both of them are able to overcome the uncertainty limit $\bar{N}^{-2}$.
This seems to contradict the BGSL \citep{Boixo2007PRL}. We note that
such a counterintuitive behavior is attributed to the problematic
definition of the BGSL. It is true as the fundamental limit for the
states of a definite photon number, but it is false for the cases
with a fluctuating photon number. A similar phenomenon has also been
observed in linear phase estimation \citep{Anisimov2010PRL,Zhong2020}.
In order to circumvent this problem, introducing a more general sensitivity
limit valid for both cases has been of interest in several studies
only related to linear phase estimation  \citep{Hofmann2009PRA,Zwierz2010PRL,Hyllus2010PRL,Pezze2015PRA},
but it is still an open question in the field of nonlinear Mach-Zehnder
interferometry. We will further discuss this problem in Sec. IV by
generalizing the BGSL into cases associating with the fluctuating
number of photons .

\section{Ultimate sensitivities determined by QCR bound}

In what follows, we wish to evaluate the ultimate sensitivities in
the above scenarios based on quantum estimation theory, which states
that, whatever measurement scheme is employed, the phase uncertainty
of an unbiased estimator $\varphi_{{\rm est}}$ is determined by quantum
Cram\'er-Rao (QCR) bound as

\begin{eqnarray}
\delta\varphi_{{\rm est}} & \geq & \frac{1}{\sqrt{\upsilon F}},\label{eq:QCBR}
\end{eqnarray}
where $F$ is the so-called quantum Fisher information (QFI). It has
been proven that the sensitivity by Eq.~\eqref{eq:error_propagation_formula}
could saturate the QCR bound with optimal measurement observables
\citep{Hotta2004PRA}, of which the condition is, however hardly satisfied
in practical applications \citep{Zhong2014JPA}. Thus it is often
desirable to seek nearly optimal measurements which could closely
approach the QCR bound. 

To identify the effect of parity measurement on our case, we need
to compare the sensitivities with parity detection as derived in the
above section to the ultimate sensitivities by Eq.~\eqref{eq:QCBR}
under the same circumstance. For simplicity, throughout this manuscript,
we assume that the system is noiseless, in the sense that quantum
states of consideration are pure. Hence, given $\rho_{{\rm in}}=\vert\psi_{{\rm in}}\rangle\langle\psi_{{\rm in}}\vert$,
the QFI in our quantum interferometry setting is given by
\begin{align}
F & =4\left[\langle\psi_{{\rm in}}\vert G^{2}\vert\psi_{{\rm in}}\rangle-\langle\psi_{{\rm in}}\vert G\vert\psi_{{\rm in}}\rangle^{2}\right],\label{eq:QFI_pure}
\end{align}
with $G\equiv B_{1}^{\dagger}(a^{\dagger}\!a)^{2}\!B_{1}$. Note that
a phase-averaging operation is required here in calculation of the
QFI due to the lack of an external reference beam in our setting \citep{Jarzyna2012PRA}.
This is because the resolution of phase shift in interferometry may
rely on coherence between states of different numbers of photon. However,
this part of the resolution is generally not measurable when additional
resources are lacking \citep{Jarzyna2012PRA}. The issue under consideration
here is related to this case as the TMSV state featuring a fluctuating
photon number. 

\begin{figure}[t]
\begin{centering}
\includegraphics[scale=0.95]{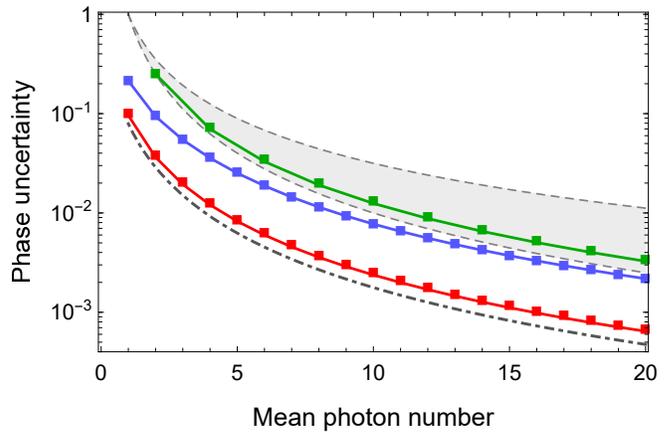}
\par\end{centering}
\centering{}\caption{(Color online) Phase uncertainties as a function of mean photon number
for TF (the green upper solid and squared-block line), TMSV (the red
lower solid and squared-block line), and EC (the blue middle solid
and squared-block line) states. The solid lines correspond to the
ultimate sensitivities limited by QCR bound and the squared-block
lines to those with parity measurement. The shaded area represents
the uncertainty region bounded by $1/\bar{N}^{3/2}$ and $1/\bar{N}^{2}$
from \citep{Boixo2007PRL}. The black dot-dashed line denotes the
sensitivity limit defined by Eq.~\eqref{eq:TMSVbound} for TMSV states\textcolor{teal}{.}\label{fig:sensitivity}}
\end{figure}

Under the phase-averaging operation, the TMSV state becomes a mixed
state that consists of a statistical ensemble of TF states, that is,
\begin{eqnarray}
\varrho_{{\rm TMSV}} & = & \sum_{n=0}^{\infty}p_{n}\left|n,n\right\rangle \left\langle n,n\right|.\label{eq:mixed_TMSVS}
\end{eqnarray}
 Although Eq.~\eqref{eq:QFI_pure} is not valid for the state of
Eq.~\eqref{eq:mixed_TMSVS}, the QFI of this state can be directly
obtained by
\begin{eqnarray}
F_{{\rm TMSV}} & = & \sum_{n=0}^{\infty}p_{n}F_{{\rm TF}}(n),\label{eq:QFI_TMSV}
\end{eqnarray}
as a consequence of the summability of the QFI \citep{Helstrom1976Book,Fujiwara2001PRA}.
Here $F_{{\rm TF}}\left(n\right)$ refers to the QFI for the TF states,
defined by Eq.~\eqref{eq:QFI_pure} by replacing $\vert\psi_{{\rm in}}\rangle$
with $\left|n,n\right\rangle $. Note that the above expression is
valid for any order of nonlinearity given in Eq.~\eqref{eq:phaseshift}.
As shown in Eq.~\eqref{eq:QFI_TMSV}, the QFI is the sum of the QFI
of the TF states with different $n$ with probability $p_{n}$, which
seems that the QFI of the TF state is essential contributing the QFI
of the TMSV state, in the sense that one can acquire the same sensitivity
reached with the TMSV state by sending a fixed number of photons in
TF states with probability $p_{n}.$ Although there is no essential
difference mathematically between the TF and TMSV states in the current
situation, the complications of preparing those states in experiments
may be far more serious. In experiments, an effective way to prepare
Fock states is to first produce pairs of light beams in the TMSV state
from a pulsed parametric down-conversion source and project one of
the beams onto a heralded Fock state by measuring another beam with
a high-efficiency photon-number-resolving detector \citep{Thekkadath2020npj}.
Obviously it is more complicated to create a heralded TF state because
of twofold equipment for creating Fock states being involved \citep{Thekkadath2020npj}.
Otherwise, it is generally a nontrivial task to produce Fock states
of large photon numbers due to low probability of multiphoton events
and low efficiency of the detector in resolving photons at high numbers
\citep{Divochiy2008NP,Sahin2013APL}. Moreover, we learn from Eq.~\eqref{eq:QFI_TMSV}
that all pairs of Fock states contained in the TMSV state contribute
to phase sensitivity. If we take the heralded TF state as the input
state, those unheralded TF states contained in the entangled resources,
which have been discarded during the state preparation, will not make
any contribution to phase sensitivity. This causes a substantial waste
of resources.

To calculate $F_{{\rm TF}}\left(n\right)$, we need to first expand
the $G$ and $G^{2}$ defined in Eq.~\eqref{eq:QFI_pure} in terms
of a multiplication of operators consisting of creation and annihilation
operators of the input modes with the help of 
\begin{eqnarray}
B_{1}^{\dagger}a^{\dagger}aB_{1} & = & \frac{1}{2}\left(a^{\dagger}a+ab^{\dagger}+a^{\dagger}b+b^{\dagger}b\right).\label{eq:BS_aa}
\end{eqnarray}
and then take the expectation over all TF states. This could be a
daunting task involving a sum of hundreds of terms to calculate. But
thanks to the orthogonality and normalization properties of Fock states,
most of these terms are vanishing except for the ones with $a$ $\left(b\right)$
and $a^{\dagger}$ $\left(b^{\dagger}\right)$ of equal count, for
instance, $\langle aa^{\dagger}a^{\dagger}ab^{\dagger}bbb^{\dagger}\rangle_{{\rm TF}}=\left(n^{2}+n\right)^{2}$
but $\langle aaa^{\dagger}ab^{\dagger}bbb^{\dagger}\rangle_{{\rm TF}}=0$.
Thus we get 
\begin{eqnarray}
\langle G\rangle_{{\rm TF}} & = & \frac{1}{2}\left(3n^{2}+n\right),\\
\langle G^{2}\rangle_{{\rm TF}} & = & \frac{1}{8}\left(35n^{4}+30n^{3}+n^{2}-2n\right).
\end{eqnarray}
Using the above expressions, it is straightforward to obtain the QFI
for Fock states: 

\begin{equation}
F_{{\rm TF}}(n)=\frac{17}{2}n^{4}+9n^{3}-\frac{n^{2}}{2}-n.\label{eq:TF_QFI}
\end{equation}
Our ultimate goal is to determine the sensitivities of Kerr phase
measurement for the TMSV states. Combing Eq.~\eqref{eq:QFI_TMSV}
with Eq.~\eqref{eq:TF_QFI} finally yields

\begin{eqnarray}
F_{{\rm TMSV}} & = & \frac{51}{4}\bar{N}^{4}+45\bar{N}^{3}+43\bar{N}^{2}+8\bar{N},\label{eq:TMSV_QFI}
\end{eqnarray}
by utilizing the following equations:
\begin{eqnarray}
\sum_{n=0}^{\infty}p_{n}n & = & \frac{\bar{N}}{2},\\
\sum_{n=0}^{\infty}p_{n}n^{2} & = & \frac{1}{2}(\bar{N}^{2}+\bar{N}),\\
\sum_{n=0}^{\infty}p_{n}n^{3} & = & \frac{1}{4}(3\bar{N}^{3}+6\bar{N}^{2}+2\bar{N}),\\
\sum_{n=0}^{\infty}p_{n}n^{4} & = & \frac{1}{2}(3\bar{N}^{4}+9\bar{N}^{3}+7\bar{N}^{2}+\bar{N}).\label{eq:meann4}
\end{eqnarray}
In addition, we also derive the accessible QFI for EC states in the
present setting,

\begin{eqnarray}
F_{{\rm EC}} & = & 2\mathcal{N}_{\alpha}^{2}\sum_{n=1}^{\infty}\left|c_{n}\right|^{2}n^{4},\label{eq:EC_QFI}
\end{eqnarray}
with $\mathcal{N}_{\alpha}=1\sqrt{2\left(1+e^{-\left|\alpha\right|^{2}}\right)}$
the normalization factor of the EC state and $c_{n}=e^{-\vert\alpha\vert^{2}/2}\alpha^{n}/\sqrt{n!}$
the corresponding superposition coefficient in terms of NOON states
of $n$ photon numbers (see Appendix A for detailed derivation).

\begin{figure}[t]
\begin{centering}
\includegraphics[scale=0.95]{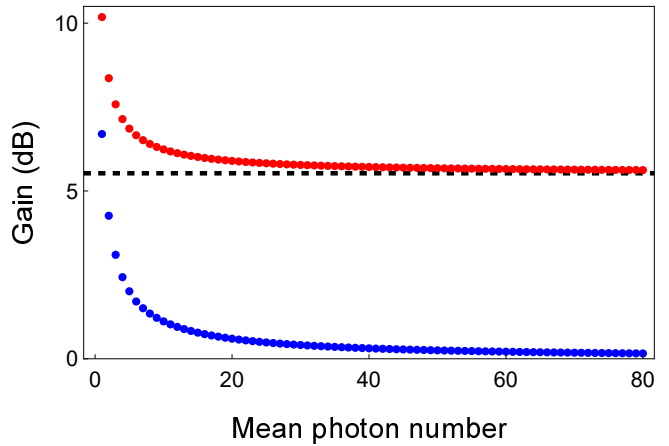}
\par\end{centering}
\centering{}\caption{(Color online) Sensitivity gain defined by Eq.~\eqref{eq:gain} for
the TMSV and EC states as a function of mean photon numbers. The red
upper dotted line corresponds to the TMSV state and the blue lower
dotted line to the EC state. The black horizontal dashed line represents
$g=-10{\rm log_{10}}\left(\sqrt{4/51}\right)\sim5.53$ dB in the infinite
$\bar{N}$ limit. \textcolor{teal}{\label{fig:gain}}}
\end{figure}

We plot in Fig.~\ref{fig:sensitivity} the phase uncertainties corresponding
to Eqs.~\eqref{eq:TF_QFI}, \eqref{eq:TMSV_QFI}, and \eqref{eq:EC_QFI}
for the three states: TF, TMSV, and EC, respectively. It is clearly
shown that the phase uncertainty achieved with parity measurement
for the EC state is identical with that determined by the QCR bound,
in the sense that parity detection is an optimal measurement for EC
states in Kerr phase estimation (see Appendix A for an explicit proof).
While they are not identical for the TF and TMSV states, the difference
between them is slightly small, in the sense that parity detection
serves as a near-optimal measurement in Kerr phase estimation with
these two states. It is also confirmed that, as suggested in the previous
section, the BGSL $1/\bar{N}^{2}$ is overcome by both the TMSV and
EC states. Unlike the EC states which lose their supersensitive advantage
as the number of mean photons becomes sufficiently large, the TMSV
states retain their capacity for overcoming the BGSL irrespective
of the photon number. 

In order to clearly show their difference, we plot in Fig.~\ref{fig:gain}
the sensitivity gain which is defined with respect to the BGSL $1/\bar{N}^{2}$
as
\begin{eqnarray}
g & \equiv & -10{\rm log}_{10}\left(\bar{N}^{2}/\sqrt{F}\right).\label{eq:gain}
\end{eqnarray}
It is clear that the behavior of the TMSV states is in sharp contrast
to the result of the EC states in that they display a supersensitive
performance only in the region of a very modest photon number and
perform equally well as the NOON states for larger $\bar{N}$ (see
Sec. IV for demonstration of NOON states being able to saturate the
BGSL of Kerr phase estimation). Similar results have been observed
in \citep{Joo2012PRA} where a common reference beam is involved.
Remarkably, the supersensitive advantage for the TMSV state is always
maintained for all $\bar{N}$ and a gain of $5.53$ dB is still expected
for a large $\bar{N}$, while there is no potential gain for the EC
state for sufficiently large $\bar{N}$.

\section{Sensitivity limits for nonlinear Mach-Zehnder interferometry}

As demonstrated in Sec. II, the supersensitive performance of the
TMSV states over the BGSL is caused by the problematic definition
of the BGSLs in the cases involving photon number fluctuation. Below,
we address this problem by introducing a more general sensitivity
limit for Kerr phase estimation with probe states of a fluctuating
number of photons.

To solve this problem, we first revisit the method applied in \citep{Boixo2007PRL}
to derive the generalized sensitivity limits for single-parameter
estimation with the $k$-order nonlinear coupling Hamiltonian. Assume
the phase accumulation is represented as a unitary operation $U_{\varphi}=\exp\left(-iH\varphi\right)$
where the generator $H$ is the coupling Hamiltonian of $N$ systems
of the form \citep{Boixo2007PRL,Napolitano2011Nat}
\begin{eqnarray}
H & = & \sum_{\left\{ i_{1},i_{2},\cdots,i_{k}\right\} }h_{i_{1}}\otimes h_{i_{2}}\otimes\cdots\otimes h_{i_{k}},\label{eq:k-coupling}
\end{eqnarray}
with the sum running over all subsets of $k$ systems and $h_{i_{k}}$
the dimensionless Hamiltonian of the $i_{k}$-th subsystem. As shown
in Eq.~\eqref{eq:QCBR}, the phase sensitivity is theoretically limited
by the inverse of the QFI, which means that the larger value of the
QFI is the higher sensitivity of phase estimation that could be acquired.
Given a $U_{\varphi}$, the QFI is upper bounded by
\begin{eqnarray}
\sqrt{F} & \leq & 2\Delta H\leq\left\Vert H\right\Vert ,\label{eq:seminorm}
\end{eqnarray}
where the first inequality is due to the fact that the QFI equals
the variance for pure states and is less than the variance for mixed
states \citep{Braunstein1996AP} and $\left\Vert H\right\Vert $ is
the operator seminorm of a Hermitian operator $H$ defined as $\left\Vert H\right\Vert =\lambda_{M}-\lambda_{m}$
with $\lambda_{M}\!\left(\lambda_{m}\right)$ the maximum (minimum)
eigenvalue of $H$ \citep{Boixo2007PRL}. These inequalities indicate
that the estimation sensitivity limit is solely determined by the
coupling Hamiltonian of the system. For the symmetric $k$-body coupling
of Eq.~\eqref{eq:k-coupling}, we have 
\begin{eqnarray}
\left\Vert H\right\Vert  & \leq & \sum_{\left\{ i_{1},i_{2},\cdots,i_{k}\right\} }\left\Vert h_{i_{1}}\otimes h_{i_{2}}\otimes\cdots\otimes h_{i_{k}}\right\Vert \nonumber \\
 & \leq & \binom{N}{k}\left\Vert h_{1}\otimes h_{2}\otimes\cdots\otimes h_{k}\right\Vert ,
\end{eqnarray}
as a result of the triangle inequality property of the seminorm. Assuming
$N\gg k$ and applying Stirling's approximation to the above expression
finally yields the sensitivity limit that scales as \citep{Boixo2007PRL,Napolitano2011Nat}
\begin{eqnarray}
\delta\varphi & \sim & \frac{k!}{N^{k}\left\Vert h_{1}\otimes h_{2}\otimes\cdots\otimes h_{k}\right\Vert }\sim\frac{1}{N^{k}}.
\end{eqnarray}
This limit was first proposed by Boixo \textit{et al}. \citep{Boixo2007PRL}.
Note that the above expression simply provides a rough sensitivity
limit for nonlinear phase estimation of a fixed number of particles
$N$, but without assuming a specific form of $H$.

Now we apply the above method to analyze the sensitivity limit in
nonlinear Mach-Zehnder interferometry. According to Eq.~\eqref{eq:phaseshift},
one can identify $H=\left(a^{\dagger}a\right)^{2}$. Note that directly
submitting this Hamiltonian into Eq.~\eqref{eq:seminorm} may obtain
an unachievable upper bound of sensitivity due to an immeasurable
global phase. To derive a more tight sensitivity bound, we resort
to the Schwinger representation as $J_{x}=\left(a^{\dagger}b+ab^{\dagger}\right)/2$,
$J_{y}=\left(a^{\dagger}b-ab^{\dagger}\right)/2i$ and $J_{z}=\left(a^{\dagger}a-b^{\dagger}b\right)/2$.
Under this representation, the Kerr Hamiltonian $H$ can be divided
into two parts as
\begin{equation}
H=\frac{\hat{N}^{2}}{4}+H_{{\rm eff}},\quad H_{{\rm eff}}=J_{z}^{2}+\hat{N}J_{z},\label{eq:phaseshift-1}
\end{equation}
with $\hat{N}=a^{\dagger}a+b^{\dagger}b$ the total photon number
operator. Consider a probe state of fixed photon number $N$ which
can be written as follows
\begin{eqnarray}
\left|\psi_{N}\right\rangle  & = & \sum_{n=0}^{N}C_{n}\left|n,N-n\right\rangle .\label{eq:pureprobe}
\end{eqnarray}
By changing into the basis space spanned by the common eigenstates
$\vert j,m\rangle$ of the operators $J^{2}=J_{x}^{2}+J_{y}^{2}+J_{z}^{2}$
and $J_{z}$, the expression of Eq.~\eqref{eq:pureprobe} can be
rewritten as $\left|\psi_{N}\right\rangle =\sum_{m=-j}^{j}C_{m}\left|j,m\right\rangle $
with $j=N/2$. After the evolution with the Hamiltonian $\hat{N}^{2}/4$
the probe state $\left|\psi_{N}\right\rangle $ remains unchanged
up to a global phase which cannot be measured. Hence the sensitivity
limit is given by maximizing $F=4\Delta^{2}H_{{\rm eff}}$ only dependent
on the effective Hamiltonian $H_{{\rm eff}}$ given in Eq.~\eqref{eq:phaseshift-1}.
An optimal probe state $\left|\psi_{N}\right\rangle $ to maximize
the variance of $H_{{\rm eff}}$ is the equally weighted superposition
of $\vert j,-j\rangle$ and $\vert j,j\rangle$ up to an arbitrary
relative phase, i.e., $\left(\vert j,-j\rangle+e^{i\phi}\vert j,j\rangle\right)/\sqrt{2}$,
where $\vert j,-j\rangle$ and $\vert j,j\rangle$ correspond to the
maximum and minimum eigenvalues of $H_{{\rm eff}}$, respectively.
It can be equivalently expressed in the Fock basis as NOON states
$\left|\psi_{{\rm NOON}}\right\rangle =\left(\vert N0\rangle+e^{i\phi}\vert0N\rangle\right)/\sqrt{2}$,
with which the QFI takes the maximum value of $F_{{\rm NOON}}=N^{4}$,
in the sense that the sensitivity limit scales as $\delta\varphi\sim1/\left(\sqrt{\upsilon}N^{2}\right)$,
which is in agreement with the BGSL for second-order nonlinear phase
estimation \citep{Boixo2007PRL}. The same sensitivity limit and optimal
probe states would be obtained if assuming $H=\hat{N}J_{z}$, which
has been theoretically proposed and experimentally studied in atomic
systems \citep{Boixo2008PRL,Napolitano2011Nat}. While it is different
for $H=J_{z}^{2}$ \citep{Kitagawa1993PRA,Luis2004PLA,Rey2007PRA,Zhong2014CPB}
which is known as the one-axis twisting Hamiltonian in the atomic
system. For this Hamiltonian, the sensitivity limit should scale with
$\delta\varphi\sim4/\left(\text{\ensuremath{\sqrt{\upsilon}}}N^{2}\right)$
and the optimal probe states for saturating the limit is $\left|\psi_{N}\right\rangle =\left(\vert j,0\rangle+e^{i\phi}\vert j,j\rangle\right)/\sqrt{2}$.

Below we relax the constraint by allowing the total particle number
to be fluctuating. Here, we simply follow the method used in \citep{Hyllus2010PRL}
to derive the generalized sensitivity limit for nonlinear phase estimation
with the variable particle number. The general states of the variable
photon number can be represented in the form of
\begin{eqnarray}
\varrho & = & \sum_{N=0}^{\infty}p_{N}\left|\psi_{N}\right\rangle \left\langle \psi_{N}\right|,\label{eq:separable_state}
\end{eqnarray}
under the assumption of absence of a suitable phase reference beam
\citep{Bartlett2007RMP,Hyllus2010PRL,Genoni2011PRL,Jarzyna2012PRA,Zhong2017PRA}.
This state can be obtained from a generic two-mode pure state $\vert\psi\rangle=\sum_{n,n^{\prime}}C_{n,n^{\prime}}\vert n,n^{\prime}\rangle$
by taking the phase-averaged operation \citep{Genoni2011PRL,Jarzyna2012PRA,Zhong2017PRA}.
Correspondingly, the state of Eq.~\eqref{eq:separable_state} is
identified with $p_{N}=\sum_{n=0}^{N}\left|C_{n,N-n}\right|^{2}$
and 
\begin{eqnarray}
\left|\psi_{N}\right\rangle  & = & \frac{1}{\sqrt{p_{N}}}\sum_{n=0}^{N}C_{n,N-n}\left|n,N-n\right\rangle .
\end{eqnarray}
In Eq.~\eqref{eq:separable_state}, it is an incoherent statistical
ensemble of pure states of the form in Eq.~\eqref{eq:pureprobe}
with a different number of photons. Based on the result derived for
Eq.~\eqref{eq:pureprobe}, the maximum QFI with respect to Eq.~\eqref{eq:separable_state}
is bounded by 
\begin{eqnarray}
F\left(\varrho\right) & = & \sum_{N}p_{N}F\left(\vert\psi_{N}\rangle\right)\leq\sum_{N}p_{N}N^{4}=\langle\hat{N}^{4}\rangle.\label{eq:QFI_mix}
\end{eqnarray}
Thus the true sensitivity limit of nonlinear Mach-Zehnder interferometry
should scale as 
\begin{equation}
\delta\varphi\sim1/\Big(\text{\ensuremath{\sqrt{\upsilon\langle\hat{N}^{4}\rangle}}}\Big),\label{eq:truelimit}
\end{equation}
when applying probe states with a fluctuating number of photons, such
as the case encountered in our study. 

Now let us come back to the initial question. In our case the input
state is a phase-averaged TMSV state $\varrho_{{\rm TMSV}}$ given
by Eq.~\eqref{eq:mixed_TMSVS}. Thus the corresponding probe state
is the state after applying the first beam splitter on the input state,
i.e., $\varrho_{{\rm TMSV}}^{b}=B_{1}\varrho_{{\rm TMSV}}B_{1}^{\dagger}$.
The expectation value of the operator $\hat{N}^{4}$ with respect
to $\varrho_{{\rm TMSV}}^{b}$ is equivalent to that with respect
to $\varrho_{{\rm TMSV}}$, i.e., $\langle\hat{N}^{4}\rangle_{\varrho_{{\rm TMSV}}^{b}}=\langle\hat{N}^{4}\rangle_{\varrho_{{\rm TMSV}}}$,
due to the commutation of $[\hat{N},B_{i}]=0\;\left(i=1,2\right)$.
It is thus straightforward to obtain
\begin{eqnarray}
\langle\hat{N}^{4}\rangle_{\varrho_{{\rm TMSV}}} & = & 24\bar{N}^{4}+72\bar{N}^{3}+56\bar{N}^{2}+8\bar{N},\label{eq:TMSVbound}
\end{eqnarray}
by using the results of $\left\langle n,n\right|\hat{N}^{4}\left|n,n\right\rangle =16n^{4}$
and Eq.~\eqref{eq:meann4}. We plot in Fig.~\ref{fig:sensitivity}
the true sensitivity limit in our situation by combing Eqs.~\eqref{eq:truelimit}
and \eqref{eq:TMSVbound}, and learn that the limit is clearly not
overcome by the TMSV states. This means that the sensitivity bound
given by Eq.~\eqref{eq:truelimit} is applicable to Kerr phase estimation
with states of fluctuating particle number, but the BGSLs fail.

\section{Conclusion}

In this paper we have analytically discussed the phase enhancement
of both TF and TMSV input states for a Kerr phase estimation using
the QFI. We have shown that the TF states can approach the BGSL proposed
by Boixo \textit{et al}. \citep{Boixo2007PRL}, while the TMSV states
can lead to a supersensitivity beyond the BGSL for any power of intensity
of incident light, which is in sharp contrast to the EC states that
display a supersensitive performance only in the region of a very
modest photon number. With high power density a sensitivity gain of
$5.53$ dB with respect to the BGSL could be still acquired for the
TMSV states. Meanwhile, on the basis of error propagation formula,
we identify parity detection as a quasioptimal measurement for both
TF and TMSV states and a genuine-optimal measurement for the EC state
in the present Kerr nonlinear phase estimation settings. 

Moreover, we elaborate that the supersensitive behavior observed with
the TMSV state is attributed to the problematic definition of the
BGSL for cases associating with a fluctuating number of photons. To
address this problem, we propose a generalized BGSL which is applicable
for these cases with probe states of a fluctuating number of photons,
to which our scheme belongs. Our work may shine some light on quantum
supersensitive measurements based on a Mach-Zehnder interferometer
with nonlinear Kerr media. 

\section*{Acknowledgments}

We are grateful to Stefan Ataman for reading and providing suggestions
and to the two anonymous referees for their enlightening comments
and suggestions for our paper. This work was supported by the NSFC
through Grants No. 12005106, the Natural Science Foundation of the
Jiangsu Higher Education Institutions of China under Grant No. 20KJB140001
and a project funded by the Priority Academic Program Development
of Jiangsu Higher Education Institutions. Y.B.S. acknowledges support
from the NSFC through Grants No. 11974189. L.Z. acknowledges support
from the NSFC through Grants No. 12175106.

\section*{Appendix A: Sensitivity reached with entangled coherent states \label{sec:AppA}}

\makeatletter \renewcommand{\theequation}{A\arabic{equation}} \makeatother \setcounter{equation}{0}

In this appendix, we calculate the phase sensitivities with EC states
in our considered case. The EC state can be understood as a superposition
of NOON states with different photons 
\begin{equation}
\vert\psi_{{\rm EC}}\rangle=\mathcal{N}_{\alpha}\sum_{n=0}^{\infty}c_{n}\left[\vert n\rangle\vert0\rangle+\vert0\rangle\vert n\rangle\right],\label{eq:ECS}
\end{equation}
with $\mathcal{N}_{\alpha}=1\sqrt{2\left(1+e^{-\left|\alpha\right|^{2}}\right)}$
the normalization factor and $c_{n}=e^{-\vert\alpha\vert^{2}/2}\alpha^{n}/\sqrt{n!}$
the superposition coefficient. This state can be generated by powering
a coherent state into one input port mode of a beam splitter and a
coherent superposition of macroscopically distinct coherent states
into another input port \citep{Luis2001PRA}. It has been demonstrated
that the EC state of small photon numbers can overcome the sensitivity
reached with NOON states \citep{Joo2012PRA}. According to the result
given by Eq.~(19) in Ref.~\citep{Joo2012PRA}, the QFI of the EC
states is approximately expressed as
\begin{eqnarray}
F_{{\rm EC}}^{r} & = & \bar{N}^{4}+10\bar{N}^{3}+13\bar{N}^{2}+2\bar{N},\label{eq:JooResult}
\end{eqnarray}
for $\left|\alpha\right|\gg1$ such that $\mathcal{N}_{\alpha}=1/\sqrt{2}$.
Obviously, the value of Eq.~\eqref{eq:TMSV_QFI} is larger than above,
in the sense that TMSV states outperform EC states in Kerr phase estimation. 

However, the sensitivity given by Eq.~\eqref{eq:JooResult} cannot
be generally saturated with rare photon-counting detection if without
introducing additional resources \citep{Molmer1997PRA,Demkowicz-Dobrzanski2009PRA,Jarzyna2012PRA},
e.g., parity measurement applied in our manuscript where the reference
beam is absent. In our case, a phase-averaged operation is required
to derive an accessible sensitivity for the EC state. After the phase-averaged
operation, the state of Eq.~\eqref{eq:ECS} is straightforwardly
expressed as
\begin{eqnarray}
\varrho_{{\rm EC}} & = & 2\mathcal{N}_{\alpha}^{2}\sum_{n=0}^{\infty}\left|c_{n}\right|^{2}\vert n::0\rangle\langle n::0\vert,\label{eq:phaseaveragedECS}
\end{eqnarray}
where we have introduced the notation $\left|n::0\right\rangle \equiv\left(\vert n\rangle\vert0\rangle+\vert0\rangle\vert n\rangle\right)/\sqrt{2}$
for simplicity. Reminding one that the QFI of NOON states is equal
to $F_{{\rm NOON}}=N^{4}$ \citep{Joo2012PRA}, as demonstrated in
Sec.~IV, the QFI for the state of Eq.~\eqref{eq:phaseaveragedECS}
can be obtained by 
\begin{eqnarray}
F_{{\rm EC}} & = & 2\mathcal{N}_{\alpha}^{2}\sum_{n=1}^{\infty}\left|c_{n}\right|^{2}F_{{\rm noon}}\!\left(n\right)=2\mathcal{N}_{\alpha}^{2}\sum_{n=1}^{\infty}\left|c_{n}\right|^{2}n^{4}.\quad\quad\label{eq:phaseaveragedQFI}
\end{eqnarray}
For larger amplitude $\left|\alpha\right|\gg1$, the expression of
Eq.~\eqref{eq:phaseaveragedQFI} approximately reduces to 
\begin{eqnarray}
F_{{\rm EC}} & = & \bar{N}^{4}+6\bar{N}^{3}+7\bar{N}^{2}+\bar{N},\label{eq:phaseaveragedQFIexc}
\end{eqnarray}
which is less than Eq.~\eqref{eq:JooResult} for the case where a
common reference beam must be established.

In what follows, let us calculate the sensitivity attained by parity
detection in the above considered scenario. Similar to the case with
the TMSV state, the expectation value of $\Pi_{b}$ for EC states
can be expressed as the weighted linear combination of the expectations
of $\Pi_{b}$ for NOON states with different photon numbers as 
\begin{eqnarray}
\langle\Pi_{b}\rangle_{{\rm EC}} & = & 2\mathcal{N}_{\alpha}^{2}\sum_{n=0}^{\infty}\left|c_{n}\right|^{2}\langle\Pi_{b}\rangle_{{\rm noon}},\label{eq:ECSparity}
\end{eqnarray}
where
\begin{eqnarray}
\langle\Pi_{b}\rangle_{{\rm noon}} & \equiv & \langle n::0\vert U_{\varphi}^{\dagger}B_{2}^{\dagger}\Pi_{b}B_{2}U_{\varphi}\vert n::0\rangle\nonumber \\
 & = & \begin{cases}
2, & n=0,\\
\cos\left(n^{2}\varphi\right), & n\neq0.
\end{cases}\label{eq:parityNOON}
\end{eqnarray}
Interestingly, with the help of Eq.~\eqref{eq:parityNOON}, we find
that
\begin{eqnarray}
\Delta\varphi & = & \frac{1}{\sqrt{\upsilon}}\frac{\sqrt{1-\langle\Pi_{b}\rangle_{{\rm NOON}}^{2}}}{\left|\frac{\partial\langle\Pi_{b}\rangle_{{\rm NOON}}}{\partial\varphi}\right|}=\frac{1}{\sqrt{\upsilon}N^{2}}.
\end{eqnarray}
This indicates that parity detection could saturate the sensitivity
limit $1/N^{2}$ independent of the true value of $\varphi$ , in
the sense that it is a global optimal measurement for Kerr phase estimation
with NOON states. A similar result has also been found in linear phase
estimation \citep{Giovannetti2006PRL,Seshadreesan2013PRA,Zhong2014JPA}.
With Eqs.~\eqref{eq:ECSparity} and \eqref{eq:parityNOON}, the sensitivity
for EC states attained by parity detection is given by 
\begin{eqnarray}
\Delta\varphi & = & \frac{1}{\sqrt{\upsilon}}\frac{\sqrt{1-\langle\Pi_{b}\rangle_{{\rm EC}}^{2}}}{\left|\frac{\partial\langle\Pi_{b}\rangle_{{\rm EC}}}{\partial\varphi}\right|}\nonumber \\
 & = & \frac{\sqrt{1-\left(2\mathcal{N}_{\alpha}^{2}\left[2\left|c_{0}\right|^{2}+\sum_{n=1}^{\infty}\left|c_{n}\right|^{2}\cos\left(n^{2}\varphi\right)\right]\right)^{2}}}{\sqrt{\upsilon}\left|2\mathcal{N}_{\alpha}^{2}\sum_{n=1}^{\infty}\left|c_{n}\right|^{2}n^{2}\sin\left(n^{2}\varphi\right)\right|}.\quad\quad\label{eq:error-ECS}
\end{eqnarray}
The above expression is explicitly simplified to $\Delta\varphi=1/\sqrt{\upsilon F_{{\rm EC}}}$
in the asymptotic limit $\varphi\rightarrow0$, in the sense that
parity detection is responsible for saturating the QCR bound for any
power intensity of incident lights. \bibliographystyle{apsrev4-1}
\bibliography{C:/ZW/manuscripts/me/ZW}

\end{document}